\documentclass[prl,footinbib,reprint,showpacs,floatfix]{revtex4-1}%
\usepackage{epsfig}
\usepackage{amsfonts}
\usepackage{amsmath}
\usepackage{amssymb}
\usepackage{graphicx}
\usepackage{bm}
\usepackage{mathptmx}  
\begin{document}
\title{Magnetism in Closed-shell Quantum Dots: Emergence of Magnetic Bipolarons}
\author{Rafa{\l} Oszwa\l dowski,$^{1,2}$ Igor \v{Z}uti\'{c},$^{1}$ and %
          A.~G.~Petukhov$^3$}

\affiliation{$^1$Department of Physics, University at Buffalo, NY 14260-1500, USA}
\affiliation{$^2$Instytut Fizyki, Uniwersytet M. Kopernika, 87-100 Toru\'{n}, Poland}
\affiliation{$^3$South Dakota School of Mines and Technology, Rapid City, SD 57701, USA}

\begin{abstract}
Similar to atoms and nuclei, semiconductor quantum dots exhibit formation
of shells. 
Predictions of magnetic behavior of the dots are often based on the shell
occupancies. Thus, closed-shell quantum dots
 are assumed to be inherently nonmagnetic. Here, we propose a possibility of 
magnetism in such dots doped with magnetic 
impurities. On the example of the system of two interacting
fermions, the simplest embodiment of the closed-shell structure, we 
demonstrate the emergence of a novel broken-symmetry ground state that is
neither spin-singlet nor spin-triplet. We propose
experimental tests of our predictions and the magnetic-dot structures to 
perform them.
\end{abstract}
\maketitle

Formation of shell structure is a ubiquitous feature in finite 
fermionic systems, such as atoms, nuclei, and
quantum dots (QDs) \cite{Hanson2007:RMP,*Reimann2002:RMP,*Maekawabook}.
An effective potential, in which the fermions are assumed to move 
independently, can be attributed to the underlying mean field, arising from an 
interplay of particle-particle interaction and 
confinement. Open-shell atoms, 
e.g., Ag and Fe, undergo a spontaneous symmetry breaking of the mean field 
and are magnetically active due to their spin-polarized ground state (GS). For 
open-shell QDs doped with transition-metal atoms, typically Mn, strong exchange
coupling between a carrier spin and the impurity spin is 
expected
\cite{Seuf2002,Besombes2004:PRL,Govo2005,Fernandez-Rossier2004:PRL,
*Abolfath2007:PRL,*Abolfath2008:PRL},
Such QDs exhibit magnetic 
ordering, which persists even up to room temperature
\cite{Beau2009,Ochsenbein2009:NT,Sellers2010:PRB}.
In contrast, closed-shell fermionic systems, e.g., noble gases, 
are known for their 
stability and the total spin-zero GS, making them magnetically
inert \footnote{We focus on ground states with zero angular momentum}.

According to a theorem by Wigner \cite{Lieb1962}, the GS
of any non-magnetic two-electron system, including a two-electron QD, 
is a spin-singlet. Thus, it would seem that closed-shell QDs doped with Mn do
not allow magnetic ordering. However, on the example of a two-particle 
(two electrons or holes)
system, we show that the Mn doping does alter the 
magnetic properties of closed-shell QDs. Surprisingly, we find a GS, 
which is neither a singlet nor a triplet, 
and allows ordering of Mn-spin,
owing to the spontaneously broken time-reversal symmetry \cite{Zhan2007}.
This mechanism of magnetism is different than in the
open-shell systems, such as bulk (Ga,Mn)As or (Cd,Mn)Te \cite{Zutic2004:RMP}. 
By definition, the open-shell systems 
have more of either "spin-up" or "spin-down" carriers. This is in contrast to 
the magnetic closed-shell state considered here, characterized by zero total
spin projection.

Carriers confined in a QD interact with magnetic ions via
contact exchange interaction, described by 
\begin{equation}
H_{ex}=-\left(J_{ex}/N_{0}\right) 
\sum_{ij,\alpha\beta}\hat{\textrm{s}}_{i\alpha
}\textrm{g}_{\alpha\beta}\hat{\textrm{S}}_{j\beta}\delta\left(  
\bm{r}_{i}-\bm{R}_{j}\right),  \label{Eq.Hex}%
\end{equation}
where $N_{0}$ is the cation density. The exchange integral, $J_{ex}$, is 
typically $\sim 0.1$~eV for electrons, and 
$\sim-1$~eV for holes. 
Carrier and magnetic ion positions are
denoted by $\bm{r}_{i}$ and $\bm{R}_{j}$ respectively;
$\hat{\textbf S}$ and $\hat{\textbf s}$ are the Mn and carrier spins.
$g$-tensor describes possible exchange-coupling anisotropy,
which is
caused by spin-orbit interaction combined with the
quasi-two-dimensional shape of QDs. 
In many semiconductors, this anisotropy is almost negligible for electrons. 
In contrast, for confined holes,
the spin-orbit coupling leads to a strong
anisotropy with ``easy axis" along the growth direction $z$
 \cite{Dorozhkin2003:PRB}.
Thus,  Eq.~(\ref{Eq.Hex}) reduces to the Ising
Hamiltonian for the heavy hole $\pm3/2$ pseudospin subspace.

We focus on a two-carrier QD, the simplest example of a closed-shell 
system. 
The total Hamiltonian, %
$H=H_{f}+H_{ex}$, 
contains the fermionic 
part $H_f$, which employs a typical two-dimensional (2D) model for
two carriers in a QD \cite{Kyriakidis2002:PRB},
$ 
H_{f}=-\hbar^{2}/\left(2m^\ast\right)\left(  \nabla_{1}^{2}+\nabla
_{2}^{2}\right)  +m^\ast\omega_{0}^2\left(  r_{1}%
^{2}+r_{2}^{2}\right)/2  +e^{2}/\left(4\pi\epsilon r_{12}\right)$%
, where $\hbar$ is the Planck constant, $m^\ast$ the effective mass,
$\omega_{0}$ determines 
the 2D confinement, $e$ is the electron charge,
$\epsilon$ the QD's dielectric constant,  and $r_{12}=|\bm{r}%
_{1}-\bm{r}_{2}|$.
 The Coulomb interaction is characterized by the effective
 Rydberg energy $Ry^{\ast}=m^\ast e^{4}/[32\left(  \pi\epsilon
 \hbar\right)  ^{2}]$.
We express the two-fermion wavefunction $\Phi$ as
\begin{equation}
\Phi=\sum_{\sigma,\sigma^{\prime}}\varphi_{\sigma\sigma^{\prime}}\left(
\bm{r}_{1},\bm{r}_{2}\right)  \chi_{\sigma}\left(  1\right)
\otimes\chi_{\sigma^{\prime}}\left(  2\right).   \label{Eq.aa}%
\end{equation}
Here $\chi_{\sigma}(1)$ and $\chi_{\sigma}(2)$ are spinors
of the carriers 1 and 2, and $\sigma=\pm1/2$ 
(or $\uparrow$, $\downarrow$) correspond
to the spin projection along the quantization axis $z$.
The Pauli principle requires $\varphi_{\sigma\sigma^{\prime}}\left(
\bm{r}_{1},\bm{r}_{2}\right)  =-\varphi_{\sigma^{\prime}\sigma}\left(
\bm{r}_{2},\bm{r}_{1}\right)  $. Thus,
$\varphi_{\uparrow\uparrow}$ and $\varphi_{\downarrow\downarrow}$
must be antisymmetric functions of $\bm{r}_{1}$ and $\bm{r}_{2}$, while
$\varphi_{\uparrow\downarrow}(\bm{r} _{1},\bm{r}_{2})$ 
and $\varphi_{\downarrow\uparrow}(\bm{r}%
_{2},\bm{r}_{1})$ transform into each other and could be neither symmetric
nor antisymmetric. %
\begin{figure}[htb]
\centering
\includegraphics[
scale=0.4,
angle=-90,
clip=true, 
viewport=0.0in 0.0in 4.8in 8.5in
]{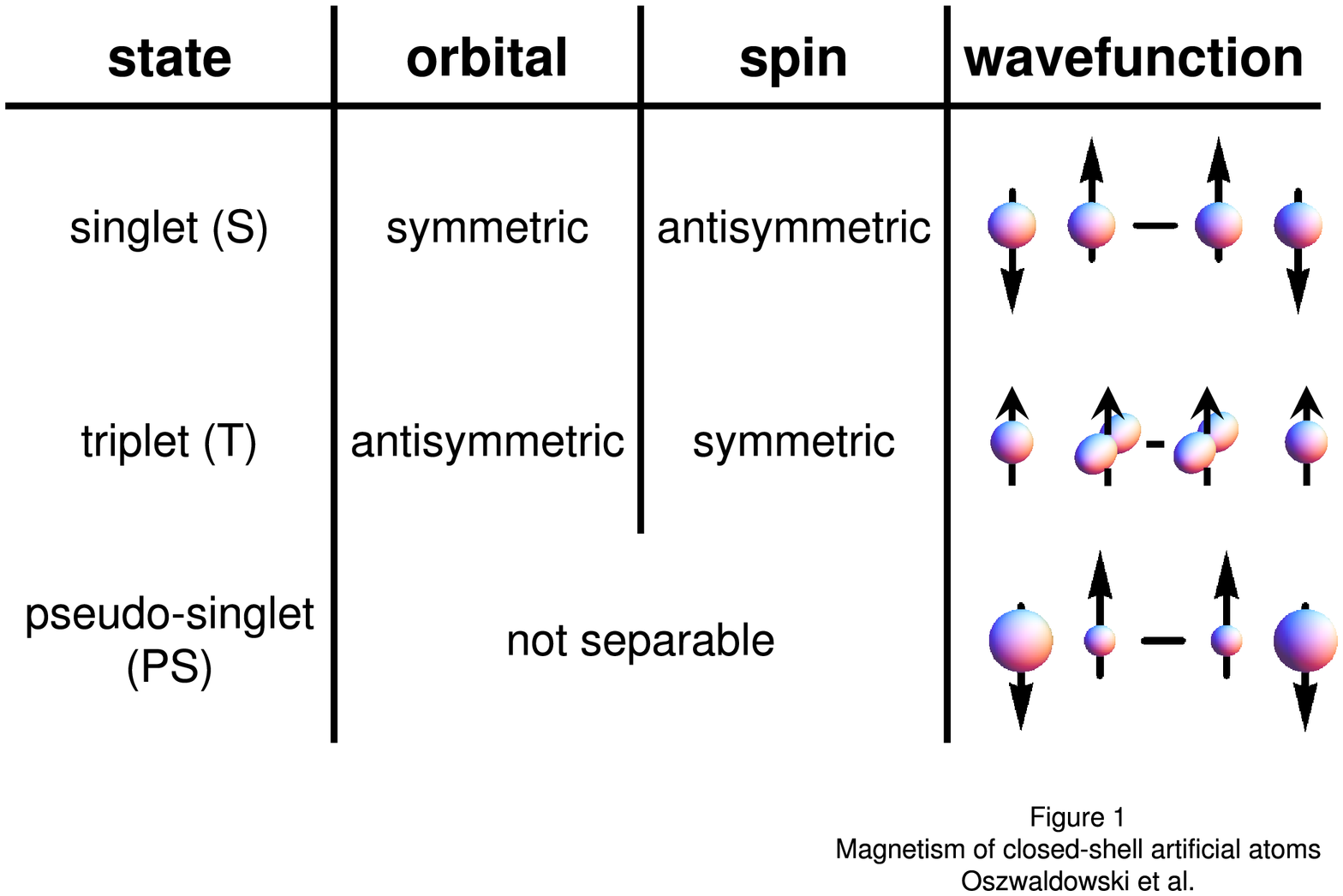}%
\caption{(color online) Two-fermion states 
described by
the symmetry of the orbital and spin part of their wavefunctions. Arrows show
spin projections (up or down), while radii of the spheres indicate
the extent of orbitals. For the S state, the radii of the orbitals
corresponding to spin-up/-down are the same. For PS, the 
spin-down orbital is larger, [Eq.\ (\ref{Eq.sing})], 
leading to finite
spin density and a magnetically-active state. 
}%
\label{Fig.symmetry}%
\end{figure}%
To understand the origin of these states, we
first consider 
the \textquotedblleft non-magnetic\textquotedblright\ 
$H_{f}$, spin-independent and invariant under the
$\bm{r}_{1}\leftrightarrow\bm{r}_{2}$ interchange. Thus, its 
eigenstates are also
eigenstates of the total carrier spin, and $\varphi_{\sigma\sigma^{\prime}%
}\left(  \bm{r}_{1},\bm{r}_{2}\right)  $ are either symmetric or
antisymmetric. The GS of any system of
two identical spin-1/2 fermions, described by a spin-independent Hamiltonian,
is a singlet with $\sigma+\sigma^{\prime}=0$ and $\varphi_{\uparrow
\downarrow}\left(  \bm{r}_{1},\bm{r}_{2}\right)  =\varphi
_{\uparrow\downarrow}\left(  \bm{r}_{2},\bm{r}_{1}\right)$
\cite{Mattisbook,Lieb1962}.
Excited states are either
singlets, or triplets with $\varphi_{\sigma\sigma^{\prime}}\left(
\bm{r}_{1},\bm{r}_{2}\right)  =-\varphi_{\sigma\sigma^{\prime}}\left(
\bm{r}_{2},\bm{r}_{1}\right)  $.
Symmetry of the singlet (ground-state) and triplet states is
illustrated in Fig.~\ref{Fig.symmetry}.
\begin{figure}[htb]
\centering
\includegraphics[
scale=0.45,
clip=true
]{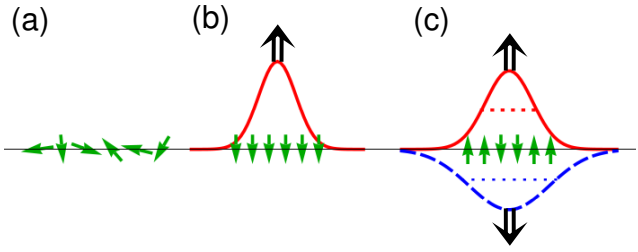}
\caption{(color online) Emergence of magnetic bipolarons in QDs. (a) Without 
carriers, Mn spins (light arrows) are randomly
oriented. Double arrows in (b, c) show carrier's
spin projection associated with the orbitals (solid  and dashed  lines). 
(b) With 1 carrier, a magnetic polaron forms, lowering the total
energy of the system, due to the coupling of the carrier's spin with the
induced Mn magnetization. Mn spins align in one direction.
(c) Two
carriers assemble in a PS state, forming what we term
a magnetic bipolaron. The
sign of the Mn-spin projection depends on the sign of the 
carrier-spin density
(difference between dashed  and solid  curves). 
The extent of the orbitals 
(length of dotted lines) is different.}%
\label{Fig.order}%
\end{figure}%

The above classification does not fully apply when $H_{ex}$ is included.
We first consider a 
two-hole system, assuming the Ising exchange, which allows
to express any eigenfunction of $H$ as
$
\psi=\Phi\left(  \left\{  \mathbf{r}_{i},s_{i}\right\}  ;
\left\{ S_{jz}\right\}  \right)  \prod_{j}\chi^{J}\left(  S_{jz}\right)$, 
 where $s_i$ are the spin variables, $S_{jz}$ is the spin projection of 
the $j$-th impurity, and $\chi^{J}(S_{jz})$ is an eigenfunction of
$\hat{\mathrm{S}}_{jz}$: $\hat{\mathrm{S}}_{jz}\chi^{J}%
(S_{jz})=S_{jz}\chi^{J}(S_{jz})$, where $S_{jz}=-J,\ldots,J$ with $J=5/2$. 
This separation of $\psi$ (or ``classical approximation" for Heisenberg spins)
is correct to order $N_\mathrm{Mn}^{-1/2}$, where $N_\mathrm{Mn}$
is the number of Mn
spins \cite{Wolff}. It has been widely used in the literature 
\cite*{Diet1983,*Wolff,%
{[{The separation is similar to that of the
Born-Oppenheimer approximation, }]{Marderbook}}}.
The separation is exact in the Ising case, so that approximations,
such as the variational method used below, are needed only for the
carrier subspace.
The two-hole $\Phi$ depends on $\left\{ S_{jz}\right\}$ 
\emph{parametrically},
and it is this coupling that leads to formation of magnetically
ordered states.
 The $z$-projection of the
total carrier spin, $\Sigma$, is a good quantum number,
so that the Hilbert space splits into three orthogonal subspaces
with values $\Sigma=\sigma+\sigma^{\prime}=$~0,~1 and $-1$
\cite*{%
[{Similar considerations apply to 2-electron na\-no\-struc\-tu\-res
placed in magnetic field, see e.g. }]
[{. In those cases the breaking of time-reversal
symmetry is not spontaneous.}]%
{Wendler1995:PRB}}.
We show that the GS is never a
singlet, i.e., $\varphi_{\sigma\sigma^{\prime}}\left(  \bm{r}%
_{1},\bm{r}_{2}\right)  \neq\varphi_{\sigma\sigma^{\prime}}\left(
\bm{r}_{2},\bm{r}_{1}\right)  $.
Instead, it is either a triplet (T) with $|\Sigma|=1$,
or what we term a pseudo-singlet (PS), with $\Sigma=0$,
which reflects its closed-shell character. 
Unlike the typical singlet, PS
leads to ordering of the magnetic moments of the open-shell 
$d$-orbitals of Mn, due to breaking of time-reversal symmetry.

Magnetic polarons form by aligning a ``cloud" of 
Mn-spins by a single
carrier localized in, e.g., a QD or in an impurity potential
\cite{Diet1983,*Wolff,Terr1992}.
The consequences of presence of two carriers in a magnetic QD
are shown in Fig.~\ref{Fig.order}. We predict 
that, even when the PS is the GS, a magnetic bipolaron is formed,
despite vanishing $\Sigma$.
\begin{figure}[htb]%
\centering
\includegraphics[
scale=0.58,
clip=true, viewport=0.6in 1.9in 6.5in 7.2in
]{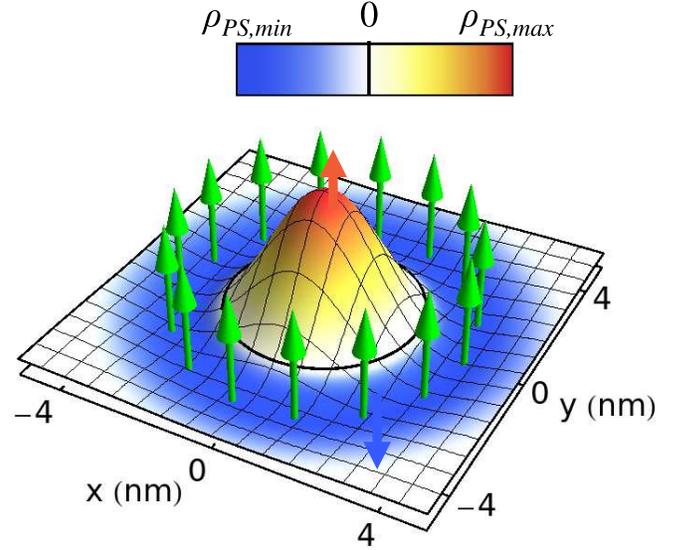} 
\caption{Spin corral. Colored surface: The hole-spin density
$\rho_{\textrm{PS}}$
(arb. units) of PS. Black circle indicates $\rho%
_{\textrm{PS}}\left(  r=R_{0}\right)  =0$. Green arrows: 
Mn spins, placed at a radius $R_{C}$, which maximizes the stability of the
ferromagnetic alignment. 
Red and blue arrows:
The more probable hole-spin projections at two positions. The
parameters: $\hbar\omega_{0}=\Delta_{0}=30$ meV, 
$Ry^{\ast}=\hbar\omega_{0}/10$,
$m^\ast =0.5m_{0}$, $w=1$ nm \cite{EPAPS:PRL:2010}. 
}%
\label{Fig.szparag}%
\end{figure}%
The finite exchange interaction is possible due to
different Bohr radii of the
\textquotedblleft up\textquotedblright\ and \textquotedblleft
down\textquotedblright\ orbitals of PS, see
the lowest row of Figs.~\ref{Fig.symmetry} and \ref{Fig.order}(c).
The non-zero exchange and symmetry breaking 
is particularly obvious 
for Mn spins arranged in a ring (\textquotedblleft spin corral\textquotedblright
\thinspace), Fig.~\ref{Fig.szparag}. These spins are
``ferromagnetically" ordered, their common direction marks one of the two
possible states (Mn spins pointing either up or down,
perpendicular to the QD plane),
corresponding to two stable magnetic-bipolaron solutions separated by an
anisotropy barrier.
The latter is defined by the strongly anisotropic hole $g$-tensor.

To analyze magnetic-bipolaron states and the symmetry breaking
induced by $H_{ex}$, we approximate the GS 
of two interacting holes 
using two alternative trial wavefunctions. The first one is the PS
$\left( \Sigma=0 \right)$%
\begin{align}
\Phi_\textrm{PS}=\frac{1}{\sqrt{2}}&\left[  u\!\left(  \bm{r}_{1}\right)
d\!\left(  \bm{r}_{2}\right)  \chi_{\uparrow}\left(  1\right)  \otimes
\chi_{\downarrow}\left(  2\right) \right. \nonumber\\
&\left.
 -u\!\left(  \bm{r}_{2}\right)
d\!\left(  \bm{r}_{1}\right)  \chi_{\uparrow}\left(  2\right)  \otimes
\chi_{\downarrow}\left(  1\right)  \right],  \label{Eq.sing}%
\end{align}
where $u,d=\sqrt{2/\pi}%
L_{u,d}^{-1}\exp\left(  -r^{2}/L_{u,d}^{2}\right)$ 
are single-carrier
orbitals corresponding to spin \textquotedblleft up\textquotedblright%
and \textquotedblleft down\textquotedblright\ respectively,
and $L_{u,d}$ are the variational parameters.
Comparing this with Eq.~(\ref{Eq.aa}), we find
$\varphi_{\uparrow\downarrow}\left(  \bm{r}%
_{1},\bm{r}_{2}\right)  =2^{-1/2}u\left(  \bm{r}_{1}\right)  d\left(
\bm{r}_{2}\right)$~\cite*{%
[{This form of $\varphi_{\uparrow\downarrow}$ and 
$\rho\!\left(  \bm{R}_{j}\right)$ 
correspond to the unrestricted Hartree-Fock approximation, see }]%
{Pielabook}}.
The second is a $\Sigma=1$ triplet%
\begin{align}
&\Phi_{\textrm T}=
\varphi_{\uparrow\uparrow}\!\left(  \bm{r}_{1},\bm{r}%
_{2}\right)  \chi_{\uparrow}(1)\otimes\chi_{\uparrow}(2),\label{Eq.trip}
\end{align}%
where
$\varphi_{\uparrow\uparrow}\!\left(  \bm{r}_{1},\bm{r}_{2}\right)
=-\varphi_{\uparrow\uparrow}\!\left(  \bm{r}_{2},\bm{r}_{1}\right)
=2/\!\left(  \pi L_{T}^{3}\right) \times$ $\left[  r_{1}e^{i\phi_{1}}-r_{2}%
e^{i\phi_{2}}\right]
e^{\left[  -\left(  r_{1}^{2}+r_{2}^{2}\right)
/L_{\textrm T}^{2}\right]}$ is the orbital part, in coordinates
$r,\ \phi$, and with one variational parameter $L_{T}$. 

The exact treatment of the isotropic $g$-tensor 
(for two-electron QDs) requires a large dimension
of the Hilbert space. The problem can be circumvented by 
replacing the Mn-spin operators with classical spin vectors 
\cite{Diet1983,*Wolff}.
Then, the two PS and two T solutions 
(Mn spins pointing up or down, see above), 
found for holes, form continua (one for PS and one for T),
corresponding to Mn spins aligned along
arbitrary directions. 
Any particular solution preserves the form given by
Eq.~(\ref{Eq.aa}) [with the $z$-axis parallel to spontaneous magnetization],
and it corresponds to formation of the magnetic bipolaron.
The distinct feature of the isotropic
case is lack of an anisotropy barrier between different solutions belonging
to the same continuum.

Owing to the disk-like shape of typical self-assembled and vertical QDs,
the $z$-dependent Schr\"odinger equation is factorized out, while
the height, $h$, (along $z$) of such QDs is usually small. This allows
to assume that only the lowest level of this equation is relevant
\footnote{We take the corresponding wavefunction to be $1/\sqrt{h}$}.
Thus, for the PS state [recall Eq.~(\ref{Eq.sing})],
the matrix element of $H_{ex}$ is 
$E_{ex}=-J_{ex}/(h N_{0})\sum_{j}\rho_\textrm{PS}\left(  \bm{R}%
_{j}\right)  S_{jz}$, where $\rho_\textrm{PS}\left(  \bm{R}_{j}\right)  =%
\left[  \left\vert u\left(  \bm{R}_{j}\right)  \right\vert
^{2}-\left\vert d\left(  \bm{R}_{j}\right)  \right\vert^{2}\right]\!/2$.
If $L_u\neq L_d$, then $E_{ex}\neq0$ in a magnetic QD.

In general: For
$H_{ex}=0$ (non-magnetic QDs), 
PS reduces to a singlet for any $Ry^\ast$, because the
non-magnetic total energy functional $E_{\textrm S}$ reaches a minimum, 
$E_{\textrm S}^{0}$, for
$L_{u}^{0}=L_{d}^{0}\equiv L_{\textrm S}^{0}$ (0 indicates a
variational minimum). In all studied systems with $H_{ex}\neq0$, however, 
$u^0\neq d^0$, i.e., PS does not reduce to a singlet.
Because of its larger spin density,
T may become the GS, despite its non-magnetic energy
being higher than $E_{\textrm S}^{0}$.
(T as the magnetic GS
was discussed by Govorov~\cite{Govo2005,Govorov2008:CRP}).
The functionals $E_{\textrm{PS},{\textrm T}}$ reach minima,
when Mn spins are antiparallel to hole-spin density.

Because of the exchange coupling, Eq.~(\ref{Eq.Hex}), the total energy
functional contains a term linear in the carrier-spin density 
$s(\bm{r})$. This term leads to instability of the closed-shell singlet 
with $s(\bm{r})=0$. PS, 
on the other hand, must satisfy a weaker, integral constraint 
$\int\!s(\bm{r})d\bm{r} = 0$. Thus, any small variation of the 
wavefunction $\Phi$ that promotes a non-zero 
$s(\bm{r})\propto\rho_\textrm{PS}$, while preserving
the integral constraint, leads to the
instability of the singlet state, because the
variations of the kinetic and the Coulomb 
energies contain only second or higher powers of $s(\bm{r})$. 

\begin{figure}[htb]
\centering
\includegraphics[
scale=0.63,
angle=-90,
clip=true, viewport=0.5in 1.5in 5.6in 7.0in
]{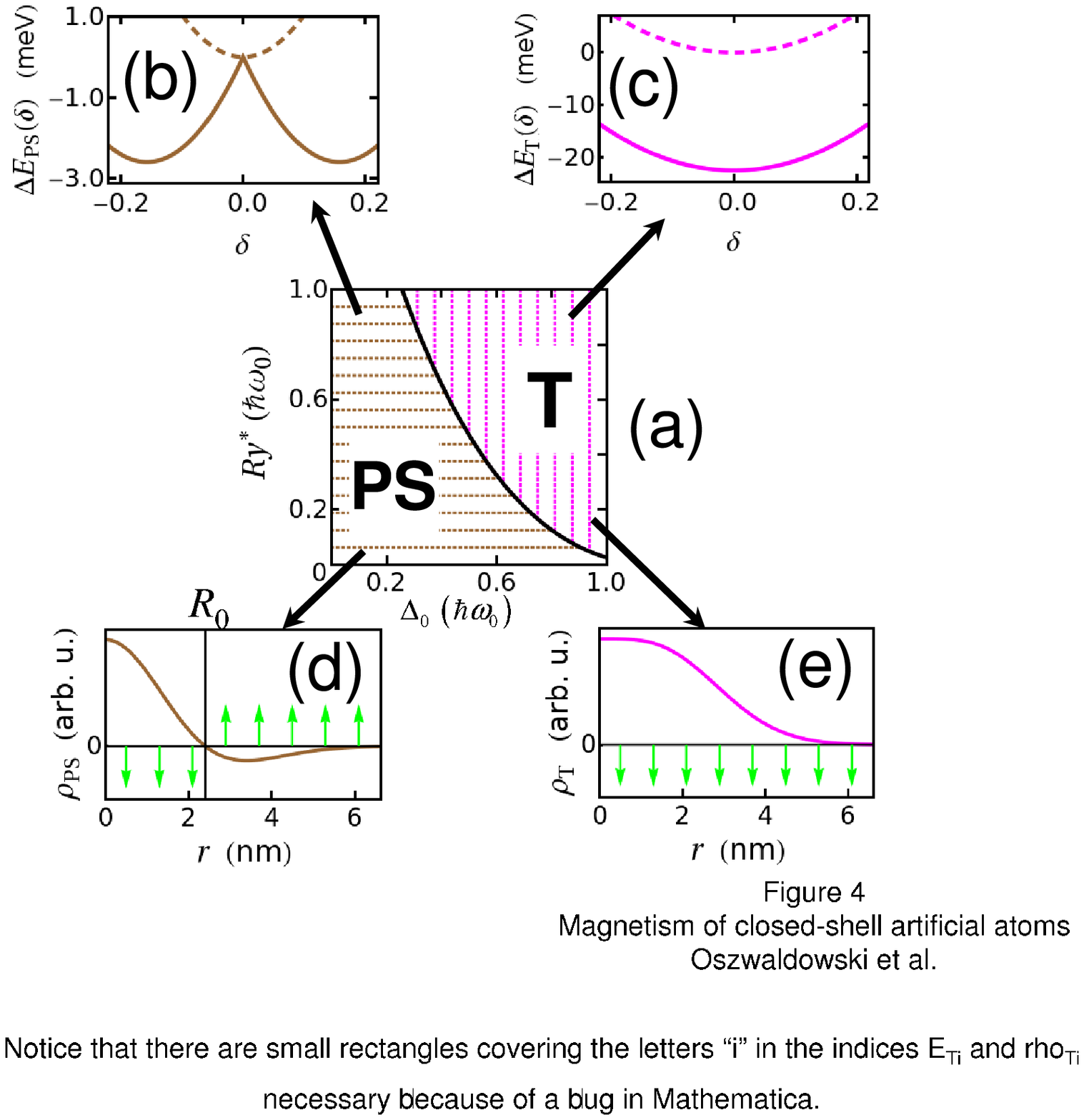} 
\caption{(color online)
(a) Phase
diagram of two-hole ground states for homogeneous
distribution of Mn. Horizontally (vertically) hatched area 
indicates the ranges
of $Ry^{\ast}$ and $\Delta_{0}$,
for which the ground state is PS (T). PS reduces
to singlet only for $\Delta_{0}=0$. (b)
for PS, and (c) for T show the non-magnetic (dashed) and total (solid)
energies as a function of $\delta$. 
Panels (d) and (e) show the hole-spin densities 
(at the corresponding variational minima), 
along any direction in the $x-y$ plane, 
with arrows indicating the Mn-spin profiles. The parameters for (b)-(e) 
are $\hbar\omega_{0}=30$ meV, $\Delta_{0}=\frac{3}{4}\hbar\omega_{0}$, 
$Ry^{\ast}=\hbar\omega_{0}/10$,
$m^\ast =0.5m_{0}$. Vertical line in (d): $R_{0}=L_{S}^{0}/\sqrt{2}$.}%
\label{Fig.phasediagram}%
\end{figure}

We now describe our results for two particular distributions of Mn spins,
 starting with homogeneous Mn content $x$. 
Figure~\ref{Fig.phasediagram}(a) shows the phase diagram
of PS and T. The former remains the
GS for moderate values of $Ry^{\ast}$ and of the saturated exchange
splitting  $\Delta_{0}=Jx\left\vert J_{ex}\right\vert $
\footnote{Figure~\ref{Fig.phasediagram}(a) also allows to determine which of 
PS, T is the GS, when the confinement strength, $\hbar\omega_0$, 
(or the related in-plane size of the QD) changes, while keeping constant 
$\Delta_0$ and $Ry^\ast$.
For this, one needs to scale the ratios $\Delta_0/\hbar\omega_0$ and 
$Ry^\ast/\hbar\omega_0$. A confinement-driven singlet-triplet
transition was noted in \cite{Govorov2008:CRP}}.
We analyze the GS
by considering small variations, $\delta$, 
of the characteristic lengths from their non-magnetic values
$L_{\textrm S}^{0}$ and $L_{\textrm{T,nm}}^{0}$.
We write $L_{u,d}=L_{\textrm S}^{0}\left(1\pm\delta\right)$ for PS, 
$L_{\textrm T}=L_{\textrm{T,nm}}^{0}\left(1+\delta\right)$ for T,
and treat $\delta$ as a variational parameter, 
see Figs.~\ref{Fig.phasediagram}(b), (c), and (d).
The quantity $\Delta E_{\textrm{PS}}\!\left(  \delta\right)  
\equiv E_{\textrm{PS}}\!\left(
\delta\right)  -E_{\textrm S}^{0}$, plotted in Fig.~\ref{Fig.phasediagram}(b),
has two minima corresponding to the two opposite Mn-magnetization profiles 
mentioned above.
The $E_\textrm{PS}^0-E_{\textrm S}^0$ gap can be an order of magnitude larger for
colloidal QDs with a few nm diameter \cite{Norb2006},
suggesting stability of PS at liquid nitrogen temperatures.
Figure~\ref{Fig.phasediagram}(c) shows that 
$\Delta E_{\textrm T}\!\left(  \delta\right)\equiv E_{\textrm T}\!\left(\delta\right)%
-E_{\textrm T}^0$ 
has $\delta^2$ dependence with a single minimum.  
Unlike PS, the T wavefunction at the
variational minimum is the same as for $H_{ex}=0$ (i.e., $\delta^{0}=0$), 
while its energy is lowered by $\Delta_{0}$.
Hole-spin densities and Mn-spin profiles
corresponding to PS and T are shown in Figs.~\ref{Fig.phasediagram}(d)
and (e).
The small variation of $\Phi$ discussed above is
$\propto\delta$. The 
singlet-PS instability manifests itself as the cusp in the 
solid line in Fig.~\ref{Fig.phasediagram}(b). 

Results of a full variational calculation \cite{EPAPS:PRL:2010},
[e.g. Fig.~\ref{Fig.phasediagram}(a)],
confirm the validity of the $\delta$ approximation.
By studying the radius at which $M_z$ changes sign, we
deduce the magnetization profile for an inhomogeneous $x$ distribution, 
such that $x\!\left( r<R_{0} \right)=0$. 
Independently of other details of the inhomogeneous distribution,
$\rho_{\textrm{PS}}$ has the same sign at all Mn sites, leading to a
"ferromagnetic" alignment in a large class of QDs.
The same alignment arises when T is the GS.
 CdSe/(Zn,Mn)Se epitaxial QDs with
Mn only at the periphery were created by intentionally introducing
Mn in the material surrounding the dot~\cite{Lee2006,Seuf2002}.
The placing of individual Mn ions with a
scanning tunnelling microscope~\cite{Kitc2006,*Hirjibehedin2006:S},
is a promising path to realize the ``spin corral".
Additionally, $x\left(r<R_0\right)=0$ could be realized
in colloidal QDs, where radial segregation of impurities
occurs during growth \cite{Norb2006}.
Such systems show strong exchange coupling \cite{Beau2009}, 
and can be controllably charged \cite{Ochsenbein2009:NT}, thus avoiding
fast Auger decay in type-I QDs
with two electron-hole pairs. 
This decay can also be suppressed
using core-shell colloidal nanocrystals, equivalent to
type-II epitaxial
QDs, due to electrons-holes separation 
\cite{Sellers2010:PRB,Klim2007}.
We estimate that for colloidal QDs with $\sim5$ nm diameter \cite{Norb2006},
the singlet-triplet splitting can be $\sim 100$ meV,
resulting in a PS ground state for a wide range of parameters.

PS existence can also be experimentally verified for homogeneous-$x$ QDs, 
e.g., as a blue shift of interband
photoluminescence with magnetic field, $B$, applied along the $z-$axis
\cite{Zutic2004:RMP}. 
With increasing $B$, all the Mn spins will tend to 
align antiparallel to it, destroying the PS magnetization profile, 
Fig.~\ref{Fig.phasediagram}(d). 
Thus, PS should increase its energy, while evolving towards the ordinary 
singlet.
This effect could be observed in type-II QDs, where the electrons 
(unlike the holes)
reside in the barrier and do not modify the physical picture.
Strong exchange coupling in these QDs is seen as
magnetic-polaron formation \cite{Sellers2010:PRB}.

Our results
can be generalized to closed-shell QDs with more carriers, and
to systems, not described by Hamiltonian $H_f$, 
typically used for self-assembled \cite{Kioseoglou2008:PRL}, vertical
\cite{Kouwenhoven2001:RPP}, or lateral \cite{Kyriakidis2002:PRB} QDs. 
Shells also form
for other confinements \cite{Reimann2002:RMP}, e.g., 
colloidal QDs can be approximately 
described by spherical or ellipsoidal potential \cite{Cantele2001:NL}.
Closed-shell ($\Sigma=0$) states appear even for asymmetrical QDs with
many carriers \cite{Reimann1999:EPJD}.
More general choices of trial functions could further
stabilize the PS energy, in some cases accompanied by lowering the symmetry
of the Mn-spin alignment.

This work was supported by DOE-BES, US ONR, AFOSR-DCT, NSF-ECCS, and CAREER.

\end{document}